\documentclass[sigconf,nonacm,9pt]{acmart}

\AtBeginDocument{%
  }

\usepackage{booktabs}
\usepackage{graphicx}

\makeatletter
\def\@authorfont{\LARGE\normalfont}
\def\@affiliationfont{\LARGE\normalfont}
\makeatother

\fancypagestyle{arxivpagestyle}{%
  \fancyhf{}%
  \fancyfoot[C]{\large\normalfont\thepage}%
}

\newcommand{\MechMemRTL}{\texorpdfstring{MechMem-RTL}{MechMem-RTL}}

\begin{document}

\title{\MechMemRTL{}: Reusing Verified Mechanism Memories for LLM-Based RTL Repair}

\author{Mingyu Cheng}
\author{Junjie Gao}
\author{Jinhua Cui}
\author{Kuncai Zhong}
\affiliation[obeypunctuation=true]{%
  \institution{College of Semiconductors, Hunan University}%
  \city{}%
  \country{China}
}

\makeatletter
\gdef\addresses{%
  \@author{Mingyu Cheng, Junjie Gao, Jinhua Cui, Kuncai Zhong\textsuperscript{*}}%
  \affiliation{obeypunctuation=true}{%
    \institution{College of Semiconductors, Hunan University, China}%
    \city{}%
    \country{}%
  }%
  \affiliation{obeypunctuation=true}{%
    \institution{{\large Email: \{mycheng, junjie1, jhcui, kczhong\}@hnu.edu.cn; \textsuperscript{*}Corresponding author}}%
    \city{}%
    \country{}%
  }%
}
\makeatother

\renewcommand{\shortauthors}{Cheng et al.}

\begin{abstract}
Large language models (LLMs) can automatically repair register-transfer-level (RTL) designs. However, fixing complex sequential logic errors requires reusing past debugging experience. Existing retrieval-augmented generation (RAG) relies on task-text similarity to provide this experience. This text-based approach often misguides the model because natural language poorly reflects cycle-level hardware execution semantics. To address this, we present \MechMemRTL{}, a repair framework that reuses verifier-confirmed repair records instead of text similarity. Each stored record strictly links trigger evidence, a diagnosed failure mechanism, a local repair action, preservation constraints, and a verification summary. For a new failure, \MechMemRTL{} injects a past record only when deterministic verifier evidence is strictly compatible with the stored trigger. Otherwise, the system uses only current verifier evidence. We evaluate \MechMemRTL{} on 48 public sequential RTL tasks across six repair models. With at most two repair attempts per task, \MechMemRTL{} successfully resolves 180 out of 288 task-model pairs, outperforming standard feedback repair (109 pairs) and task-similarity RAG (107 pairs).
\end{abstract}



\maketitle
\pagestyle{arxivpagestyle}

\section{Introduction}
\label{sec:introduction}

Large language models (LLMs) have emerged as practical tools for register-transfer-level (RTL) design~\cite{verilogeval,rtllm,verigen,rtlcoder}. However, because their initial outputs rarely pass all verification checks on the first attempt, automated RTL workflows rely heavily on an iterative repair loop. Within this loop, the model proposes localized revisions based on compiler or simulator feedback. These revisions are highly constrained, as a valid repair must strictly preserve the original module interface and match cycle-accurate behavior under the target testbench.

Satisfying these cycle-accurate constraints makes fixing sequential logic particularly difficult. A single functional mismatch can stem from flawed finite-state machine (FSM) transitions, incorrect event priorities, or misaligned protocol boundaries. Current verifiers usually report the immediate failing symptom, such as a signal mismatch at a specific clock cycle. They do not identify the structural root cause. Since LLMs do not directly model hardware semantics, they struggle to map these surface symptoms to the correct code edit. Closing this gap requires the repair system to reuse prior debugging experience rather than relying solely on the latest error message.

A common approach to providing this past experience is retrieval-augmented generation (RAG) based on task-text similarity~\cite{rag,rag_benchmark}. However, in sequential RTL debugging, textual similarity is a weak proxy for hardware execution behavior. Tasks with completely different descriptions can fail due to the same state-update error. Conversely, similar task descriptions can fail for entirely different cycle-level reasons. Therefore, reusing repair experience based merely on text similarity often misguides the LLM. 

To address this limitation, we present \MechMemRTL{}, an automated framework that extracts and reuses verifier-confirmed repair mechanisms. Upon a successful repair, the system constructs a structured mechanism memory record. Each record explicitly links trigger evidence, a diagnosed failure mechanism, a localized repair action, preservation constraints, and a verification summary. For subsequent repairs, a deterministic evidence filter injects a stored record into the prompt only when the current compiler or simulator diagnostics are strictly compatible with the stored trigger. Otherwise, the context is constructed from current verifier evidence.

Specifically, we make the following contributions.

\begin{itemize}
  \setlength{\itemsep}{1pt}
  \setlength{\parsep}{0pt}
  \setlength{\topsep}{2pt}
  \item We introduce a verifier-confirmed mechanism memory designed for sequential RTL repair. It formally encapsulates five key elements: verification trigger evidence, a diagnosed failure mechanism, a targeted repair action, behavior preservation constraints, and a verification summary (see Section~\ref{sec:method}).
  \item We design a hardware-aware deterministic evidence filter. It selects past records using strict compiler and simulator diagnostics rather than task-text similarity or language model heuristics, ensuring high-fidelity guidance (see Section~\ref{sec:method}).
  \item We curate SeqRTL, an evaluation set of 48 public sequential RTL tasks with failing initial designs. We thoroughly compare our approach against standard feedback repair and task-similarity RAG across six different repair models (see Sections~\ref{sec:experiment-setup} and~\ref{sec:results}).
\end{itemize}

Experimental results confirm the effectiveness of \MechMemRTL{}. Given a strict limit of two repair attempts per task, our method successfully resolves 180 out of 288 task-model pairs. Under identical conditions, standard feedback repair and task-similarity RAG resolve only 109 and 107 pairs, respectively. We will make our framework code and evaluation artifacts publicly available.

\begin{figure*}[t]
  \centering
  \includegraphics[width=0.92\textwidth]{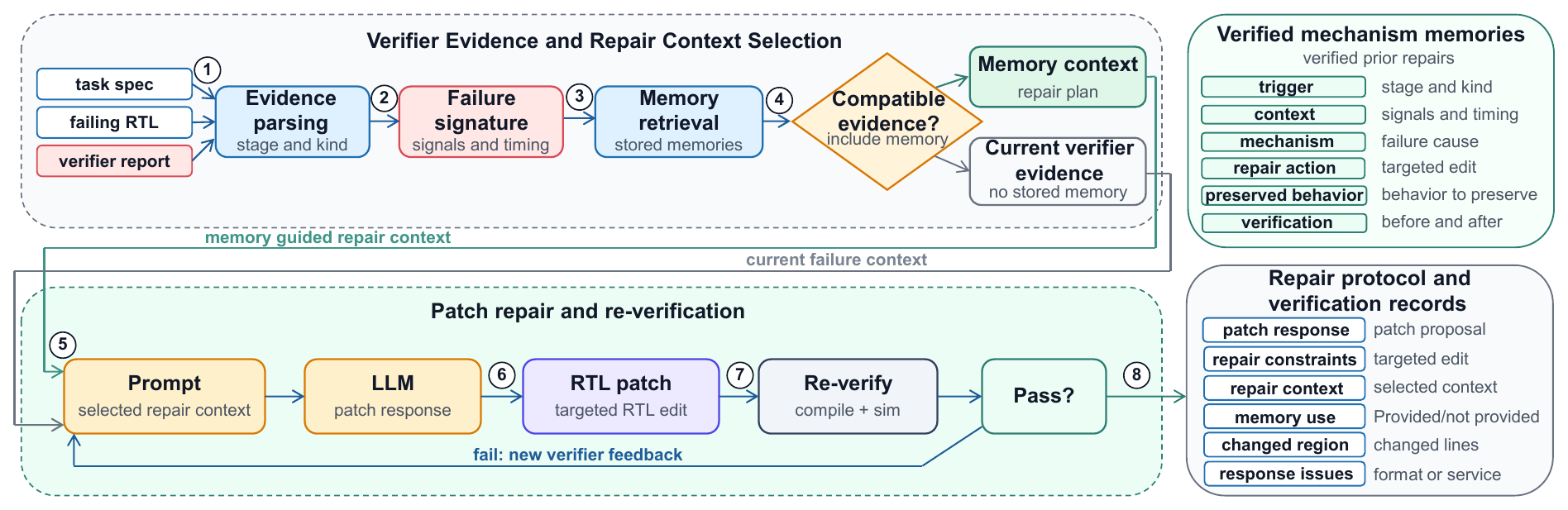}
  \caption{Workflow of \MechMemRTL{}. Verifier evidence determines whether a stored repair record guides the next repair.}
  \Description{Workflow diagram showing verifier evidence parsing, failure signature extraction, memory retrieval, evidence matching, selected repair context construction, patch repair, re-verification, and diagnostic records.}
  \label{fig:mechmem-workflow}
\end{figure*}

\section{Background and Motivation}
\label{sec:motivation}

In this section, we explain why automated RTL repair requires reusable knowledge grounded in verifier evidence rather than task text.

\textbf{RTL generation and evaluation.}\
Recent studies establish a standard paradigm for LLM-aided RTL design using executable checks~\cite{verilogeval,rtllm,autosilicon,veriagent,verilogeval_revisit,rtlbench}. In this setting, a model proposes an RTL candidate, and an automated harness verifies its functional correctness. Although domain-specific models trained on RTL data improve initial generation quality~\cite{verigen,rtlcoder}, complex designs rarely pass all checks immediately. Because many candidates fail on the first attempt, the design workflow naturally enters a repair stage.

\textbf{RTL repair and failure mechanisms.}\
Current RTL repair systems guide LLM edits using compiler errors or simulator feedback~\cite{rtlfixer,meic,hdldebugger,context_verilog_repair}, often validated against executable testbenches~\cite{fixbench_rtl,autoverifix}. However, verifier feedback typically reports a surface symptom rather than the structural root cause. For example, a simulator may flag an output mismatch at a specific clock cycle, while the actual fix requires changing an FSM transition, an event priority rule, or a protocol boundary. Recent error analyses confirm that effective repair must distinguish specific failure mechanisms rather than just binary pass or fail outcomes~\cite{rtlerrors}. This gap between immediate symptoms and structural root causes makes it necessary to reuse prior debugging experience.

\textbf{Limitations of task-similarity RAG.}\
A common approach to reusing past experience is task-similarity RAG~\cite{rag,rag_benchmark}. For sequential RTL debugging, however, text similarity is a weak proxy for hardware execution behavior. Tasks with completely different descriptions can fail due to the same state-update error, while similar descriptions can fail for entirely different cycle-level reasons. Retrieval based solely on task wording often introduces context that looks relevant but mismatches the current failure mechanism. Therefore, the repair context should be selected using evidence tied directly to compiler diagnostics, timing, and signal behavior.

\textbf{Reusable verification knowledge.}\
Executable RTL checks naturally provide this deterministic evidence. Because RTL repair is strictly constrained by module interfaces and cycle-accurate behavior, a successful repair is much more than a simple text solution. It is a verifiable debugging episode. General software repair often relies on execution traces or case-based reasoning~\cite{cbr,apr_plm,impact_apr,chatrepair,thinkrepair,hybrid_apr,experepair}. In contrast, sequential RTL debugging demands reusable artifacts strictly tied to hardware timing and signal evidence~\cite{uvllm}. \MechMemRTL{} applies this principle by capturing each successful repair as a verifier-confirmed mechanism memory record. Each record links the failure trigger, the diagnosed mechanism, the local repair action, preservation constraints, and a verification summary. We formalize the structure and filtering of these mechanism memories in the next section.

\section{Proposed Method}
\label{sec:method}

In this section, we formalize the \MechMemRTL{} framework. We first detail how the system creates and reuses verifier-confirmed repair records. We then formally define the memory record structure, explain the deterministic evidence filter that controls reuse, and describe the final patch-level diagnostics.

\subsection{Core Workflow}
\label{subsec:repair-workflow}

Figure~\ref{fig:mechmem-workflow} illustrates the core workflow of \MechMemRTL{}. The system operates in two distinct phases. These are memory creation and memory-guided repair.

During memory creation, the system extracts knowledge from successful debugging episodes. A memory record is stored only when a proposed RTL patch passes the executable testbench. This strict rule keeps the memory bank anchored in verified hardware evidence rather than unverified model hypotheses.

During memory-guided repair, the inputs consist of the task specification, the current faulty RTL candidate, and the latest verifier report. \MechMemRTL{} normalizes this verifier evidence and retrieves initial candidate records from the memory bank. A deterministic evidence filter then evaluates whether any retrieved record is strictly compatible with the current failure. If a compatible record survives this filter, it is injected into the prompt as the selected repair context. If no compatible memory record is selected, the repair context is constructed from the current verifier evidence without reusing stored memory. The model then proposes a localized patch, and the verifier evaluates the modified RTL. If the candidate still fails, the process repeats using updated verifier feedback until it reaches the predefined iteration limit.

Crucially, this workflow explicitly targets sequential RTL failures involving control and timing behavior. Such failures include finite-state machine transitions, output timing delays, protocol boundaries, counter thresholds, pulse alignments, reset logic, and state-history updates. Compile and tool-compatibility errors are also addressed when they surface within these repair loops. This bounded scope ensures that memory reuse is driven entirely by cycle-level hardware semantics.

\subsection{Mechanism Memory Record}
\label{subsec:memory-entry}

At the core of this workflow is the mechanism memory, a structured diagnostic record backed by executable verification. We formalize it as a five-element tuple \( m = (T, M, R, P, V) \). Here, \( T \) represents the trigger evidence. \( M \) is the diagnosed failure mechanism. \( R \) specifies the localized repair action. \( P \) outlines the behavior preservation constraints. Finally, \( V \) provides the verification summary.

The trigger \( T \) captures normalized compiler or simulator diagnostics. It includes the verification stage, failure kind, affected outputs, and first mismatch time if available. The mechanism \( M \) summarizes the hardware failure cause, such as a state-output timing delay or a serial bit-order inversion. The repair action \( R \) records the accepted localized code edit. The preservation field \( P \) identifies surrounding logic, interface constraints, or timing behavior that must remain unchanged. The verification summary \( V \) simply logs the transition from a failing outcome to a passing one.

While the mechanism (\( M \)) and preservation (\( P \)) fields are generated during the repair attempt, they are never stored in isolation. \MechMemRTL{} commits the complete five-element record to the memory bank only after the associated patch passes the executable testbench. This verification-gated process ensures that the reusable rationale is tied to a concrete and verifiable edit.

\begin{table}[htbt] 
\caption{Example verifier-confirmed repair record.}
\label{tab:memory-example}
\centering
\footnotesize
\setlength{\tabcolsep}{3pt}
\renewcommand{\arraystretch}{1.08}
\begin{tabular}{@{}p{0.30\columnwidth}p{0.63\columnwidth}@{}}
\toprule
\textbf{Field} & \textbf{Stored content} \\
\midrule
\textbf{Prior task} & VerilogEval shift-count task \\
\textbf{Trigger evidence} & Output \texttt{q}: 1886/2071 mismatched samples; first mismatch at time 10 \\
\textbf{Mechanism} & Wrong direction in MSB-first serial shift \\
\textbf{Repair action} & Replace \texttt{\{data, q[3:1]\}} with \texttt{\{q[2:0], data\}} \\
\textbf{Preserved behavior} & Preserve counter behavior, interface, and hold behavior \\
\textbf{Verification} & Same testbench: mismatch before repair, pass after repair \\
\bottomrule
\end{tabular}
\end{table}

Table~\ref{tab:memory-example} provides an example of a verifier-confirmed repair record. The trigger and verification fields anchor the memory in objective hardware evidence. Meanwhile, the mechanism, repair action, and preservation fields specify exactly what knowledge can be reused. Consequently, the reusable unit is not a mere copy of a previous solution. Instead, it serves as a focused repair context centered on one specific failure mechanism.

\subsection{Evidence Filtering}
\label{subsec:evidence-routing}

\begin{figure}[htbt] 
  \centering
  \includegraphics[width=\columnwidth]{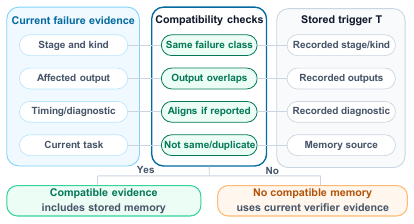}
  \caption{Deterministic evidence filtering. Matched evidence exposes memory; otherwise, repair uses current evidence.}
  \Description{Comparison diagram showing current failure evidence, deterministic compatibility checks, stored trigger evidence, and the compatible or fallback outcomes.}
  \label{fig:evidence-filtering}
\end{figure}

With the memory bank populated by these structured records, the next critical challenge is accurately identifying the correct repair context for a newly encountered failure. As illustrated in Figure~\ref{fig:evidence-filtering}, this selection process operates in two stages. First, a retrieval module proposes initial candidate records based on broad normalized verifier evidence. Second, the deterministic evidence filter rigorously screens these candidates to guarantee strict hardware compatibility with the current failure.

Instead of relying on probabilistic language model judgments, the filter extracts deterministic hardware features to perform these strict compatibility checks. These features correspond to the parallel comparison lanes in Figure~\ref{fig:evidence-filtering}, mapping the current failure evidence directly to the stored trigger \( T \). The filter systematically evaluates the verification stage, failure kind, affected output signals, and temporal diagnostic signatures. To prevent data leakage, it utilizes task metadata to exclude records originating from the same task or known duplicate tasks.

Compatibility requires a strict set of mechanism-specific constraints. For example, a compile-stage memory can only be reused for a compatible compile-stage diagnostic. A one-cycle output-timing memory requires a matching simulation mismatch pattern on a related output port or timing phase. Similarly, a serial bit-order memory requires matching signal and phase evidence. These deterministic constraints are much stricter than natural language text similarity. They prevent weakly related hardware contexts from bypassing the filter and misleading the repair model.

After filtering, \MechMemRTL{} constructs the final repair context. If compatible memories survive, the prompt includes the surviving records and requests a targeted patch. If no compatible memories survive, the retrieved records are withheld from the model, and the repair context is constructed from the current verifier evidence without reusing stored memory. Section~\ref{sec:results} reports these context-selection decisions as explicit repair diagnostics.

\subsection{Repair Patches and Diagnostics}
\label{subsec:patch-logging}

Having determined the appropriate repair context, the framework constructs the final prompt. This prompt integrates the current faulty RTL, the immediate verifier feedback, and any memory records that survive the filter. Regardless of the specific context type, the system explicitly instructs the model to diagnose the failure mechanism and return a localized patch. Furthermore, the prompt requires the model to identify the modified code region and specify which surrounding behaviors must remain functionally intact. The expected response is a targeted edit rather than a full-module rewrite.

Once the language model generates this candidate patch, the system must enforce strict hardware boundaries. The repaired RTL must strictly preserve the original module interface and keep logic modifications strictly confined to the diagnosed failure zone. To monitor adherence to these structural constraints, the framework continuously logs key process metrics. These tracked diagnostics include response validity, repair-constraint violations, full-module rewrite attempts, line modification counts, and memory filtering statistics. While these metrics do not replace the primary verified pass rate, they provide essential transparency. Ultimately, this tracking ensures that the automated debugging process strictly adheres to practical hardware engineering standards.

\section{Experimental Setup}
\label{sec:experiment-setup}

\begin{figure}[htbt] 
  \centering
  \includegraphics[width=\columnwidth]{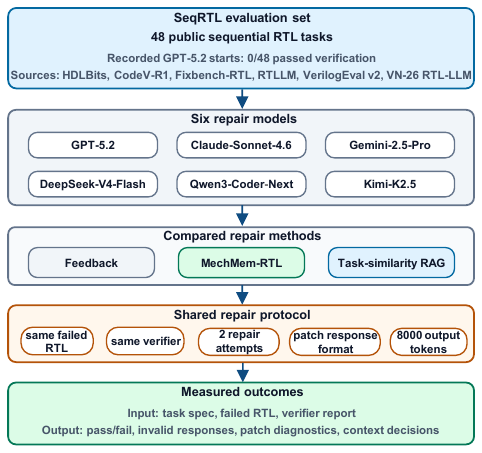}
  \caption{Matched repair protocol for SeqRTL across three repair methods and six repair models.}
  \Description{Single-column protocol diagram showing public task sources, the initial-fail GPT-5.2 repair subset, six matched repair models, three compared repair methods, shared repair controls, and measured outcomes and diagnostics.}
  \label{fig:matched-eval-protocol}
\end{figure}

To evaluate the proposed framework, we establish a strictly matched repair protocol. By fixing the task set, initial failing designs, iteration limits, and verification harness across all methods, we ensure a fair and fully reproducible comparison~\cite{verilogeval,rtllm,fixbench_rtl,swebench}. Figure~\ref{fig:matched-eval-protocol} outlines this shared evaluation pipeline.

\subsection{SeqRTL Evaluation Set}

Because our focus is automated debugging rather than initial generation, we construct SeqRTL, a dedicated evaluation set of 48 public sequential RTL tasks. To guarantee a zero initial pass rate, each task begins with a recorded faulty design originally generated by GPT-5.2. We include a task only if it features sequential control logic, has public provenance, and includes a verified reference design.

To prevent data leakage, we rigorously exclude any tasks overlapping with the source pool used to build our mechanism memory bank. We also filter out known duplicates, initially passing designs, and tasks without runnable testbenches. The final 48 tasks and their starting failures remain strictly frozen throughout the evaluation. These tasks originate from established benchmarks (e.g., HDLBits, CodeV-R1, RTLLM)~\cite{rtllm,fixbench_rtl} and cover diverse hardware behaviors, including protocol handshaking, reset initialization, edge capture, counter thresholds, and state-history updates.

\subsection{Repair Methods}
\label{subsec:repair-methods}

We compare three repair methods under a matched design. All methods share the same task specification, current faulty RTL candidate, and verifier feedback. The only difference is the additional repair context.

\textbf{Standard feedback repair.} This baseline relies exclusively on the current faulty RTL candidate and the immediate compiler or simulator diagnostic report.

\textbf{Task-similarity RAG.} This baseline represents the conventional textual retrieval approach. It retrieves up to three solved task examples based on lexical overlap across task titles, specifications, and module interfaces~\cite{rag,rag_benchmark}. To construct the prompt, these retrieved examples provide task descriptions, reference interfaces, and retrieval metadata. The target task and its known duplicates are strictly excluded from the retrieval pool to ensure fairness.

\textbf{\MechMemRTL{}.} Our proposed method utilizes a separate bank of 53 verifier-confirmed mechanism memory records, prepared independently from the evaluation runs. To maintain the integrity of this bank, a source repair episode contributes a record only when its patch successfully passes verification. During the evaluation runs, the system injects a stored record into the prompt only if the current diagnostic report is strictly compatible with the stored trigger evidence. If no compatible memory record is selected, the repair context is constructed from the current verifier evidence without reusing stored memory.

To ensure strict fairness, all methods repair the same 48 starting designs. Each method receives a maximum of two repair attempts per task and an 8000-token output limit. All methods use identical patch response formats and share the same retry rules for handling invalid formats. Importantly, models are explicitly constrained to preserve the original module interface, avoid full-module rewrites, and restrict edits strictly to the diagnosed failure zone.

Every repair attempt utilizes the identical Icarus Verilog simulation harness. The models receive text-based compiler diagnostics and testbench mismatch summaries, but no raw Value Change Dump (VCD) traces or formal counterexamples. We evaluate these methods across six repair models (GPT-5.2, Claude-Sonnet-4.6, Gemini-2.5-Pro, DeepSeek-V4-Flash, Qwen3-Coder-Next, and Kimi-K2.5), yielding a total of 864 matched evaluation runs.

\subsection{Metrics}
\label{subsec:metrics}

The primary evaluation metric is the verified repair pass rate, which measures the fraction of initially failing designs that successfully pass the verification harness within the allowed two attempts. We report the absolute repair gains of \MechMemRTL{} compared to both baselines across all six models.

\begin{figure*}[htbt!] 
  \centering
  \includegraphics[width=0.98\textwidth]{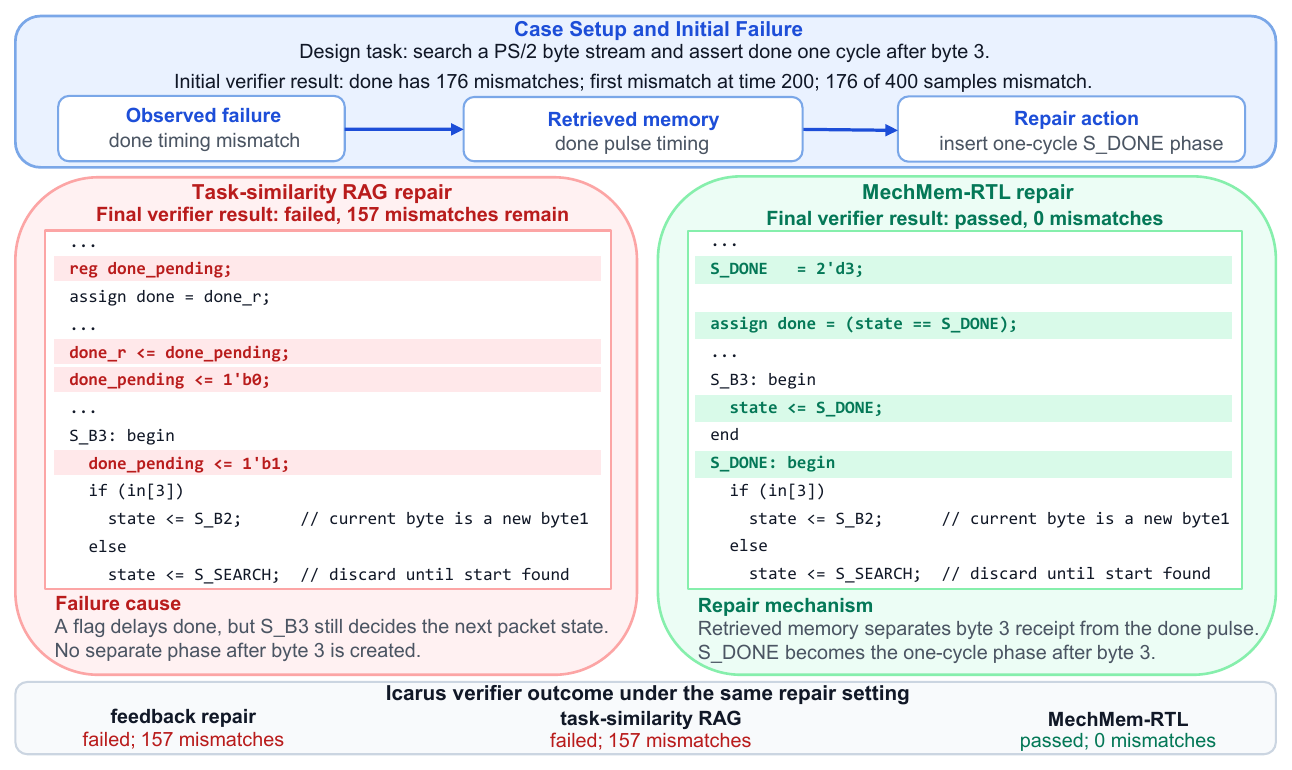}
  \caption{PS/2 packet-boundary case study with a separate one-cycle done phase.}
  \Description{Two-column case-study figure showing an initial PS/2 done timing failure, a retrieved memory, a patch action, a failed task-similarity RAG repair, a passed MechMem-RTL repair, and matched verifier outcomes.}
  \label{fig:ps2-case-study}
\end{figure*}

To further interpret the final outcomes, we monitor patch-contract diagnostics. As established in Section~\ref{subsec:patch-logging}, these track invalid response formats, line modification counts, full-module rewrite attempts, and repair-constraint violations. While these diagnostics do not replace the primary pass rate, they provide essential transparency into system stability and verify that the models adhere to practical hardware constraints under identical testing conditions.

\section{Results and Discussion}
\label{sec:results}

This section evaluates repair performance across 864 matched runs on SeqRTL. As established in Section~\ref{sec:experiment-setup}, all comparisons share identical starting designs, verifiers, repair limits, and token budgets. Invalid repair responses are strictly counted as failed attempts. Unless stated otherwise, all counts refer to task-model pairs. Table~\ref{tab:main-results} reports the aggregate repair outcomes, and Figure~\ref{fig:ps2-case-study} details a specific packet-boundary repair case.

\begin{table}[htbt!] 
\caption{Verified repair counts under the matched protocol. Gain is over the stronger baseline for each repair model.}
\label{tab:main-results}
\centering
\footnotesize
\setlength{\tabcolsep}{2.4pt}
\begin{tabular*}{\columnwidth}{@{}l@{\extracolsep{\fill}}cccc@{}}
\toprule
Repair model & Feedback & \shortstack{Task-sim.\\RAG} & \MechMemRTL{} & Gain \\
\midrule
GPT-5.2 & 22 & 23 & \textbf{36} & +13 \\
Claude-Sonnet-4.6 & 23 & 21 & \textbf{33} & +10 \\
Gemini-2.5-Pro & 19 & 19 & \textbf{26} & +7 \\
DeepSeek-V4-Flash & 25 & 22 & \textbf{29} & +4 \\
Qwen3-Coder-Next & 8 & 5 & \textbf{23} & +15 \\
Kimi-K2.5 & 12 & 17 & \textbf{33} & +16 \\
\midrule
Total & 109/288 & 107/288 & \textbf{180/288} & +65 \\
\bottomrule
\end{tabular*}
\end{table}

\subsection{Matched Repair Results}
\label{subsec:matched-repair-results}

Table~\ref{tab:main-results} summarizes the verified repairs under the shared protocol. \MechMemRTL{} achieves the highest verified repair count across all six repair models, successfully resolving 180 out of 288 task-model pairs. In comparison, standard feedback repair solves 109 pairs, and task-similarity RAG solves 107 pairs. This yields an aggregate gain of 65 repairs over the stronger baselines in the six per-model comparisons.

Notably, task-similarity RAG fails to improve upon standard feedback repair in aggregate. This validates our core motivation: natural language task text alone is an unreliable proxy for sequential hardware behavior. Retrieved examples often match at the specification level but fundamentally mismatch the current cycle-level failure, thereby misleading the model~\cite{irrelevant_context,lost_middle}.

\begin{figure}[htbt!] 
  \centering
  \includegraphics[width=0.91\columnwidth]{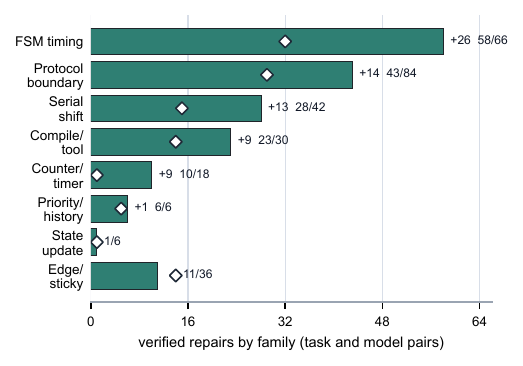}
  \caption{Verified repairs categorized by mechanism family. Bars represent \MechMemRTL{} while diamonds indicate the stronger baseline.}
  \Description{Horizontal bar chart showing verified repairs by mechanism family. Bars show MechMem-RTL repair counts, white diamonds show the stronger of the feedback and task-similarity RAG counts, and right-side labels show positive MechMem-RTL gain when present plus the MechMem-RTL repaired/total count.}
  \label{fig:family-gain}
\end{figure}

Figure~\ref{fig:ps2-case-study} illustrates this specific failure mode on a PS/2 packet-boundary task. Starting from the same faulty design, task-similarity RAG merely attempts to delay a \texttt{done} flag, leaving 157 simulation mismatches unresolved. In contrast, \MechMemRTL{} retrieves a compatible pulse-timing memory. This deterministic context guides the model to structurally introduce a separate one-cycle \texttt{S\_DONE} phase, completely resolving the mismatch.

Figure~\ref{fig:family-gain} categorizes these verified repairs by mechanism family. The most substantial gains emerge in FSM timing, protocol-boundary, and serial-shift categories. These failures demand precise adjustments to cycle-level timing and state transitions, confirming that verifier-confirmed memories provide critical structural context. Compile/tool and counter/timer families also exhibit consistent gains. Performance on state-update pairs remains tied. Interestingly, standard feedback repair resolves slightly more edge or sticky-capture pairs. Such localized capture failures typically expose sufficient diagnostic information in the immediate verifier report, making historical context less vital. This family-level breakdown comprehensively clarifies the source of our aggregate gains.

\subsection{Paired Outcomes}

\begin{figure}[htbt] 
  \centering
  \includegraphics[width=0.89\columnwidth]{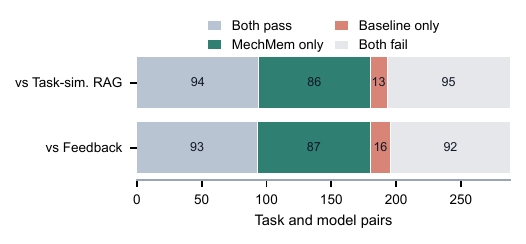}
  \caption{Paired outcomes over 288 task-model pairs. Each bar compares \MechMemRTL{} with one baseline.}
  \Description{Stacked horizontal bars showing paired outcomes versus feedback repair and versus task-similarity RAG. The bars separate task and model pairs where both methods pass, only MechMem-RTL passes, only the baseline passes, or both fail on the same pair.}
  \label{fig:paired-outcomes}
\end{figure}

To confirm that the aggregate gain represents a true capability expansion rather than a simple shift in solved tasks, Figure~\ref{fig:paired-outcomes} details the uniquely resolved pairs. Against standard feedback repair, \MechMemRTL{} uniquely resolves 87 complex failures, while missing only 16 pairs handled by the baseline. Against task-similarity RAG, the corresponding unique counts are 86 pairs for \MechMemRTL{} and only 13 pairs for the RAG baseline. These paired outcomes demonstrate that \MechMemRTL{} consistently resolves complex structural failures that remain out of reach for both baselines.

\subsection{Repair Diagnostics}

To maintain end-to-end transparency under the shared protocol, we track response format failures and count invalid outputs as failed attempts~\cite{chatrepair,thinkrepair}. Format failures are unevenly distributed. Gemini-2.5-Pro produces 61 invalid responses (mostly empty), and DeepSeek-V4-Flash produces 21 failures (often due to token limits). Other models produce at most 8 invalid responses. These metrics reveal how different model architectures interact with identical structural repair contexts.

Patch-contract diagnostics further reveal the nature of the generated fixes. On average, \MechMemRTL{} modifies 7.54 lines per task-model pair, slightly higher than the 6.38 and 6.55 lines modified by the two baselines. In sequential logic, this slight increase suggests that \MechMemRTL{} performs necessary structural refactoring rather than trivial signal toggling. Importantly, this structural awareness does not degrade overall stability: full-module rewrite attempts remain extremely rare (3 for \MechMemRTL{}, 4 for each baseline). Furthermore, \MechMemRTL{} triggers fewer repair-constraint violations than both baselines, confirming its strict adherence to hardware boundaries.

\subsection{Use of Stored Memories}

A critical design principle of \MechMemRTL{} is to reuse stored memory only when the current hardware evidence is strictly compatible with the stored trigger. Table~\ref{tab:route-summary} tracks these deterministic routing decisions.

\begin{table}[htbt!] 
\centering
\caption{Context selection and outcomes for \MechMemRTL{}.}
\label{tab:route-summary}
\begin{tabular}{lrrr}
\toprule
Repair context & Pairs & Pass & Rate \\
\midrule
With stored memory & 162 & 118 & 72.8\% \\
Without stored memory & 126 & 62 & 49.2\% \\
\midrule
Total & 288 & 180 & 62.5\% \\
\bottomrule
\end{tabular}
\end{table}

Across the six repair models, 162 task-model pairs receive a prompt augmented with stored memory, yielding 118 verified passes. The remaining 126 task-model pairs use repair contexts constructed from the current verifier evidence without stored memory, yielding 62 verified passes. This selective routing shows that \MechMemRTL{} uses historical context only when the current evidence is compatible with a stored trigger, reducing the risk of misleading the repair model with irrelevant context.

Overall, \MechMemRTL{} achieves the highest verified repair count across all six repair models under the matched protocol. These results support the central claim that sequential RTL repair benefits from evidence-based context selection rather than task-text similarity alone.

\section{Conclusion}
\label{sec:conclusion}

This paper presents \MechMemRTL{}, a framework for reusing verifier-confirmed experience in automated RTL repair. Instead of relying on unreliable text similarity, the system stores successful debugging episodes as structured mechanism memories. It reuses a memory only when the current verifier evidence is strictly compatible with the stored trigger. Across 48 public sequential RTL tasks and six repair models, \MechMemRTL{} successfully resolves 180 out of 288 task-model pairs, outperforming standard feedback repair (109) and task-similarity RAG (107). These results demonstrate that sequential RTL repair benefits from selective memory reuse grounded in deterministic verification evidence. This approach is particularly effective for resolving complex cycle-level functional errors, where task text alone provides weak and misleading guidance.

\bibliographystyle{ACM-Reference-Format-citeorder}
\bibliography{refs}

\end{document}